\documentclass[twocolumn,showpacs,prb,superscriptaddress]{revtex4}%
\usepackage{amssymb}
\usepackage{graphicx}
\usepackage{bm}%
\usepackage{amsmath}%
\usepackage{amsfonts}

\begin{document}


\title{First order $0$ - $\pi$ phase transitions in
superconductor/ferromagnet/superconductor trilayers}

\author{A.~V.~Samokhvalov}
\affiliation{Institute for Physics of Microstructures, Russian
Academy of Sciences, \\ 603950 Nizhny Novgorod, GSP-105, Russia}
\affiliation{Lobachevsky State University of Nizhny Novgorod,
Nizhny Novgorod 603950, Russia}
\author{A.~I.~Buzdin} \affiliation{Institut Universitaire
de France and University of Bordeaux, LOMA UMR-CNRS 5798, F-33405
Talence Cedex, France}

\date{\today}

\begin{abstract}

We study the thermodynamics of the diffusive SFS trilayer composed
of thin superconductor (S) and ferromagnet (F) layers. On the
basis the self-consistent solutions of nonlinear Usadel equations
in the F and S layers we obtain the Ginzburg--Landau expansion and
compute the condensation free energy and entropy of the $0$ (even)
and $\pi$ (odd) order parameter configurations. The first order
$0-\pi$ transition as a function of temperature $T$ occurs, which
is responsible for a jump of the averaged magnetic field
penetration depth $\lambda(T)$ recently observed on experiments
[N.Pompeo, et. al., Phys. Rev. B \textbf{90}, 064510 (2014)]. The
generalized Ginzburg-Landau functional was proposed to describe
SFS trilayer for arbitrary phase difference between the
superconducting order parameters in the S layers. The temperature
dependence of the SFS Josephson junction critical current
demonstrates the strong anharmonicity of the corresponding
current--phase relation in the vicinity of the $0-\pi$ transition.
In rf SQUID, coexistence of stable and metastable $0$ and $\pi$
states provides integer and half--integer fluxoid configurations.
%

\end{abstract}


\maketitle


\section{Introduction}


The ground state of the superconductor-ferromagnet-superconductor
(SFS) trilayer at zero current can be $0$ or $\pi$ state,
depending on the value of the phase difference between the
superconducting order parameters in the two S electrodes. This
phenomenon is related to the damped oscillatory behavior of the
Cooper pair wave function in the ferromagnet due to the proximity
effect \cite{Buzdin-Bulaevskii-JETPL82,Buzdin-Kuprianov-JETPL91}
(for more references and reviews, see
Refs.~\onlinecite{Buzdin-RMP05,Golubov-RMP04,Bergeret-RMP05}).
Usually experiments directed towards the observation of the
$0-\pi$ crossover in SFS trilayer were concentrated on the
measurements of the critical Josephson current $I_c$ of the SFS
junction
\cite{Ryazanov-PRL01,Kontos-PRL02,Frolov-Ryazanov-PRB04,Oboznov-Ryazanov-Buzdin-PRL06}.
The $0-\pi$ transition manifests itself in the vanishing of $I_c$
if higher--order harmonics of the current--phase relation are
negligible
\cite{Frolov-Ryazanov-PRB06,Buzdin-PRB05_0-pi-trans,Houzet-PRB05}.

Recently, an unusual electromagnetic response of SF systems was
reported as a manifestation of the Cooper pair wavefunction
oscillations inside the ferromagnet. Measurements of the London
penetration depth in thin $\mathrm{Nb/Ni}$ bilayers
\cite{Lemberger-APL08} reveal a slightly nonmonotonic dependence
of the penetration depth on the F layer thickness, which was in
accordance with the theoretical analysis \cite{Houzet-PRB09}.
Anomalous Meissner effect in hybrid SF structures was the subject
of several theoretical works
\cite{Bergeret-PRB01,Asano-PRL11,Yokoyama-PRL11} predicting an
unusual paramagnetic response of such systems. Vanishing or
inversion of the Meissner effect is believed to be attributed to
spin-triplet superconducting correlations
\cite{Bergeret-PRL01,Kadigrobov-Shekhter-EL01} generated in
inhomogeneous F layer due to proximity effect and should result in
the in-plane Fulde-Ferrell-Larkin-Ovchinnikov (FFLO) instability
\cite{FF-PR64,LO-ZETF64,Muronov-PRL12}. Unusual drop of the
screening with decrease of temperature was observed recently in
\cite{Pompeo-Samokhvalov-PRB14} by microwave measurements of the
London penetration depth $\lambda$ in
$\mathrm{Nb/Pd_{0.84}Ni_{0.16}/Nb}$ trilayers.

The transition temperature of SF structures into the normal state
has been examined both theoretically
\cite{Buzdin-Kuprianov-JETPL90_Tc-sf-multylayers,Radovic-Buzdin-PRB91_Tc-sf-multylayers,Baladie-Buzdin-PRB03_Tc-FSF,Tollis-PRB04_type-I}
and experimentally
\cite{Jiang-PRL95_Oscill-Tc,Muhge-PRL96_Oscill-Tc,Shelukhin-PRB06_pi-shift,Garifullin-Tagirov-PRB02_reentrant-Tc}
(see
Ref.~\onlinecite{Buzdin-RMP05,Sidorenko-Tagirov-AnnPhys03_Tc-bilayer},
for reviews). However, the study of the thermodynamic properties
of the phase transition between $0$ and $\pi$ states of SF hybrids
is more sparse. A first-order $0-\pi$ transition was predicted for
diffusive SFS junctions with a homogeneous F barrier, if the
current--phase relation takes into account the second harmonic
contribution \cite{Buzdin-PRB05_0-pi-trans}. Experimental evidence
of a $0-\pi$ transition in SFS ($\mathrm{Nb / Cu_x Ni_{1-x} /
Nb}$) Josephson junction was obtained from the measurements of the
temperature dependence of the critical current in
Refs.~\onlinecite{Frolov-Ryazanov-PRB04,Oboznov-Ryazanov-Buzdin-PRL06,Sellier-PRL04}.
Note that a small modulation of the thickness of the barrier may
favor the continuous $0-\pi$ transition
\cite{Buzdin-PRB05_0-pi-trans}. A first-order transitions between
$0$ and $\pi$ states by temperature variation were demonstrated in
both the clean \cite{Barsic-PRB06_type-I,Barsic-PRB07_type-I} and
dirty \cite{Tollis-PRB04_type-I} limits using numerical
self-consistent solutions of the microscopic Bogoliubov-de-Gennes
\cite{Bogolubov-JETP58,deGennes1966} or Usadel \cite{Usadel-prl70}
equations, respectively. The $0-\pi$ transition in ballistic SF
systems has been shown to have a pronounced effect on the
distribution of the Cooper pair wavefunction in the F region and
the amplitude of order parameter $\Delta(T)$ in S--layers
\cite{Barsic-PRB06_type-I,Barsic-PRB07_type-I}. The theoretical
model Proposed in Ref.~\onlinecite{Pompeo-Samokhvalov-PRB14}
argued that the observed jump $\lambda(T)$ in
$\mathrm{Nb/Pd_{0.84}Ni_{0.16}/Nb}$ trilayers is related to the
first order phase transition from $0$ to $\pi$ state on cooling.

In this work we develop a theoretical approach based on the
nonlinear Usadel equations providing a general description of
diffusive SFS junction with thin superconducting layers at the
transition from $0$ to $\pi$ state. The leakage of the Cooper
pairs weakens the superconductivity near the interface with a F
metal due to the proximity effect \cite{Buzdin-RMP05}. The
magnitude of the superconducting order parameter suppression
depends on the parameters characterizing the system such as the SF
interface transparency, the thickness of the S and F layers, etc.
For large interface transparency this effect seems to be
especially strong and results in suppression of the
superconducting order parameter and the transition temperature
$T_c$ of a thin superconducting layer in contact with a
ferromagnet metal. If the thickness of a superconducting layer is
smaller than a critical one, the proximity effect totally destroys
a superconductivity. From the self-consistent solutions we obtain
the Ginzburg--Landau expansion and compute the condensation free
energy and entropy of the possible order parameter configurations
as a function of temperature $T$. As $T$ varies, we find that the
first order phase transition between $0$ and $\pi$ states occurs,
which is responsible for a jump of the averaged penetration depth
$\lambda(T)$ observed in
Ref.~\onlinecite{Pompeo-Samokhvalov-PRB14}. We also calculate the
current--phase relation $I(\varphi)$ of the SFS junction which
reveals strong contribution of the higher harmonic terms. The
$0-\pi$ states coexistence and switching leads to new modes of
magnetic flux penetration in superconducting loop containing the
SFS junction.

The paper is organized as follows. In Sec.~\ref{ModelSection} we
briefly discuss the basic equations. We analyze the case of thin S
layers and obtain the approximate solutions of the non-linear
Usadel equations in F layer near the superconducting critical
temperature $T \lesssim T_c$. In Sec.~\ref{CritTemper} we find the
temperature  $T_c^{0,\pi}$ of the second--order superconducting
phase transition to the normal state for $0$ and $\pi$
order-parameter configurations. The Sec.~\ref{PhaseTrans} is
devoted to the analysis of the temperature-driven transition
between $0$ and $\pi$ states. We obtain the Ginzburg--Landau
expansion and find the ground states of SFS junction near the
critical temperatures $T_c^{0,\pi}$. In Sec.~\ref{CurrPhaseRelat}
we generalize the Ginzburg--Landau description for arbitrary phase
difference $\varphi$ between the superconducting order parameters
of the S layers and find the strongly nonsinusoidal current--phase
relation of the SFS junction in the vicinity of $0-\pi$
transition.
In Sec.~\ref{SingleJuncLoop} we show that
the coexistence of $0$ and $\pi$ states leads to peculiarities of
the magnetic flux penetration in superconducting loop with a
single SFS junction. Sec.~\ref{Summary} contains a brief summary
and discussion.

\section{ Model and basic equations}\label{ModelSection}

\begin{figure}[ptb]
\includegraphics[width=0.35\textwidth]{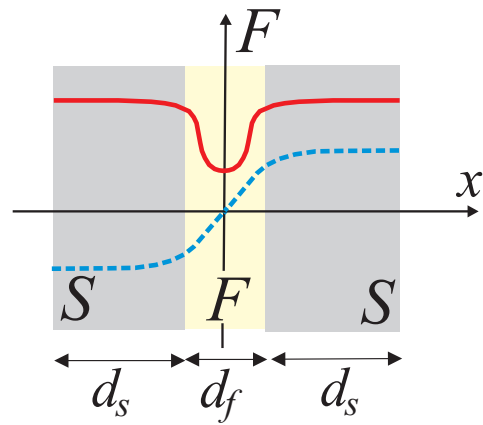}
\caption{(Color online) The schematic behavior of the pair wave
function $F(x)=F_{s,f}$ inside the SFS trilayers. The red solid
line represents approximately the behavior of the pair wave
function in an even mode ($0$-state). Due to symmetry the
derivative $\partial_{x}F_{f}$ is zero at the center of F layer.
The pair wave function in the odd mode (blue dashed line) vanishes
at the center of F layer,  and $F(x)$ has a $\pi-$shift in
diametrically
opposite points ($\pi-$state).}%
\label{Fig-1}%
\end{figure}

Let us consider a SFS trilayers with a transparent SF interfaces
and thin S layers of thickness $d_{s} \sim \xi_{s}$, where
$\xi_{s}$ is the superconducting coherence length. The considered
geometry of the SFS structure is presented in Fig.~\ref{Fig-1}.
%
%
In the previous observations of $0$-$\pi$ transition in SFS
junctions by temperature variation
\cite{Ryazanov-PRL01,Oboznov-Ryazanov-Buzdin-PRL06} the
superconducting electrodes were rather thick to overcome the
pair--breaking proximity effect of F layer and then the influence
of the $0$-$\pi$ transition on the superconducting order parameter
$\Delta(T)$ in the electrodes was negligible. From the theoretical
point of view, weak depairing in S electrodes means that the pair
potential at the SF interfaces is equal to its bulk value (the
so-called rigid--boundary condition \cite{Golubov-RMP04}). Here we
study the SFS structure with relatively thin S layers and
demonstrate that the $0$-$\pi$ transition leads to the jump of the
amplitude of the superconducting order parameter, providing the
anomalous temperature behavior of the effective penetration depth
of the whole structure.

To elucidate our results we start the qualitative discussion of
the proximity effect on the properties of SFS sandwiches if the
thickness of S layers is small enough. Due to the damped
oscillations of the pair wave function $F$ in the ferromagnetic
layer, two different order--parameter configurations are possible
inside the SFS trilayers (see Fig.~\ref{Fig-1}). The first one
corresponds to the case than the order parameter is an even
function of the coordinate $x$, chosen perpendicular to the
layers, and does not change its sign in F layer. It means that in
the ground state the superconducting phase in both S layers must
be the same ($0-$phase). For the second case the pair wave
function is odd in $x$ and cross zero at the center of the F layer
which causes a $\pi-$shift in the superconducting phase of
different S layers. This configuration corresponds to the
$\pi-$phase of the SFS structure. Since pair--breaking proximity
effect depends on the structure of the Copper pairs wavefunction
in ferromagnetic layer, a suppression of superconductivity in thin
S layers is expected to be different for $0$ and $\pi$ states. As
a result the equilibrium superconducting gaps $\Delta_{0,\pi}$ for
$0$ and $\pi$ states are different ($\Delta_{0}(T) \ne
\Delta_{\pi}(T)$), and this leads to the different effective
penetration depth $\lambda(T) \sim 1/\Delta(T)$ in $0$ and $\pi$
states. So, $0$-$\pi$ transition in SFS trilayers has to be
accompanied by a jump of $\lambda$. The coexistence of stable and
metastable states in the vicinity of $0-\pi$ transition leads to a
strong anharmonicity of the current-phase relation in the SFS
junction under consideration. As a result, peculiarities of the
magnetic flux penetration in a mesoscopic superconducting loop
containing the SFS junction are expected.

Taking in mind that the F interlayer is dilute ferromagnetic
alloys like $\mathrm{Cu_x Ni_{1-x}}$ or $\mathrm{Pd_{x} Ni_{1-x}}$
we use the Usadel equations \cite{Usadel-prl70} which are
convenient in the diffusive limit (see \cite{Buzdin-RMP05} for
details). Moreover it is important to take into account the
magnetic disorder which is already present in the magnetic alloys
and provides the main mechanism of the temperature induced
$0$-$\pi$ transition
\cite{Oboznov-Ryazanov-Buzdin-PRL06,Faure-prb06}. We describe it
by introducing the magnetic scattering time $\tau_{s}$. Note that
in practice the exchange field $h$ acting on the electron's spins
in the ferromagnet and magnetic scattering rate $\tau_{s}^{-1}$
are much larger than the superconducting critical temperature
$T_c$: $h,\, \tau_{s}^{-1} \gg T_c$. The complete nonlinear Usadel
equations for the normal $G(x,\omega,h)$ and anomalous
$F(x,\omega,h)$ Green's functions in F layer are
\cite{Oboznov-Ryazanov-Buzdin-PRL06,Faure-prb06}:
\begin{eqnarray}\label{eq:1}
    -\frac{D_f}{2} \left[ G(x,\omega,h)\, \partial_x^2 F(x,\omega,h)
        -F(x,\omega,h)\, \partial_x^2 G(x,\omega,h) \right] \nonumber \\
        +\left[ \omega + i h + \frac{G(x,\omega,h)}{\tau_s} \right]
        F(x,\omega,h) = 0\, , \qquad
\end{eqnarray}
\begin{equation}\label{eq:2}
     G^2(x,\omega,h) + F(x,\omega,h) F^+(x,\omega,h) = 1\,.
\end{equation}
Here $D_f$ is the diffusion constant in the ferromagnet, $\omega =
2\pi T (n+1/2)$ is a Matsubara frequency at the temperature $T$.
The equation for the function $F^+(x,\omega,h) = F^*(x,\omega,-h)$
coincides with Eqn.~(\ref{eq:1}) \cite{Buzdin-RMP05}. For the
$0-$state we should choose the even anomalous Green's functions
$F$ while for the $\pi-$state it should be the odd one. Using the
usual parametrization of the normal and anomalous Green functions
$G_s=\cos\theta_s$ and $F_s=\sin\theta_s$, the complete nonlinear
Usadel equation in superconucting layers can be written for
$\omega > 0$ as \cite{Buzdin-RMP05}
\begin{equation}\label{eq:3}
-\frac{D_s}{2} \partial_x^2 \theta_s
 + \omega \sin\theta_s = \Delta \cos\theta_s \,,
\end{equation}
where $D_s$ is the diffusion constant in the superconductor.
Assuming the SF interfaces to be transparent we have at $x = \pm
d_{f}/2$ \cite{Kupriyanov-JETP88}:
\begin{equation}\label{eq:4}%
    F_{s}=F,\quad\sigma_{s}\,\partial_{x}F_{s}
    =\sigma_{f}\,\partial_{x} F\,,
\end{equation}
where $\sigma_{f}$ and $\sigma_{s}$ are the normal--state
conductivities of the F and S metals, respectively. The boundary
condition at the outer surfaces $x = \pm (d_s + d_f/2)$ is
\begin{equation}\label{eq:5}
     \partial_x F_s = 0\,.
\end{equation}
For thin S-layers $d_{s}\lesssim\xi_{s}=\sqrt{D_{s}/2\pi T_{c0}}$,
the inverse proximity effect is substantial, and the Usadel
equation (\ref{eq:3}) for the S layers have to be completed by the
self-consistency equation for the superconducting order parameter
$\Delta(x)$:
\begin{equation}\label{eq:6}
    \Delta(x) = \pi T \rho \sum_{\omega} F_s(x,\omega) ,
\end{equation}
where $\rho$ is BCS coupling constant and $T_{c0}$ is the critical
temperature of a bulk sample of the material S.

For $0$ and $\pi$ states of SFS trilayers we have $F^+(x,\omega,h)
= F^*(x,\omega,-h) = F(x,\omega,h)$, and one can replace
$F^+(x,\omega,h)$ by $F(x,\omega,h)$ in Eqn.~(\ref{eq:2}). Just
below the critical temperature $T_c$ anomalous Green's functions
are small and the condition (\ref{eq:2}) can be rewritten as
\begin{equation}\label{eq:7}
     G(x,\omega,h) \simeq 1 - F^2(x,\omega,h) / 2\,.
\end{equation}

For $d_s \ll \xi_s$, where $\xi_s=\sqrt{D_s/2\pi T_{c0}}$ is the
superconducting coherence length, the variations of the functions
$\theta_s(x)$ and $\Delta(x)$ in the superconducting layers are
small: $\theta_s(x)\simeq \theta_s$, $\Delta(x)\simeq \Delta$.
Therefore, we can average Eq.~(\ref{eq:3}) over the thickness of
the S layers, using the boundary condition (\ref{eq:5}). Finally,
we obtain the following boundary condition:
\begin{equation}\label{eq:8}
    \frac{\partial\theta_s}{\partial s}{\bigg\vert_{s_f}} =
    \frac{d_s \xi_f}{\xi_s^2} \left( \frac{\Delta \cos\theta_s
    - \omega \sin\theta_s}{\pi T_{c0}} \right) \,,
\end{equation}
\begin{equation}\label{eq:9}
    \frac{\partial\theta_s}{\partial s}{\bigg\vert_{-s_f}} =
    \mp \frac{\partial\theta_s}{\partial s}{\bigg\vert_{s_f}}\,,
\end{equation}
where $s = x / \xi_f$ and $s_f = d_f / 2 \xi_f$. The top (bottom)
sign in (\ref{eq:9}) corresponds to the $0-$phase ($\pi-$phase),
respectively.

Applying the method \cite{Buzdin-PRB05_0-pi-trans} we can find the
approximate solution of the non-linear Usadel equation
(\ref{eq:1}) in F layer near the superconducting critical
temperature $T \lesssim T_c$. For the $0-$state we should choose
the even anomalous Green's functions $F$ while for the $\pi-$state
it should be the odd one:
\begin{eqnarray}\label{eq:10}
    &&F(s,\omega) \simeq
        \begin{cases}
            a \cosh(q s) \\
            b \sinh(q s)
        \end{cases} - \\
    &&\quad \frac{1}{8 k^2} \left(\alpha + \frac{3 i}{4}\right)
        \begin{cases} a^3 \cosh(3 k s), & 0-\mathrm{phase}\\
                      b^3 \sinh(3 k s), & \pi-\mathrm{phase}
        \end{cases}\, ,  \nonumber
\end{eqnarray}
where $\xi_f^2 = D_f / h$, $k^2 = 2 (\omega/h + i +
\alpha\,\mathrm{sgn}(\omega))$ and $\alpha = 1 / \tau_s h$ is the
dimensionless magnetic scattering rate. The complex wave vector
$q$ is determined by the relations:
\begin{equation}
    q^2 = k^2 \mp a^2 \left(\alpha + i / 4 \right) \label{eq:11}
\end{equation}
for $0$ and $\pi$ phases, respectively. If $T_c < \tau_s^{-1},\,
h$, we may neglect the Matsubara frequencies in the
Eq.~(\ref{eq:1}) assuming that $k^2 = (k_1 + i k_2)^2 = 2 (\alpha
+ i)$ for $\omega
>0$ :
\begin{equation}
    k_1 = \sqrt{\sqrt{1+\alpha^2}+\alpha}\,, \quad
    k_2 = \sqrt{\sqrt{1+\alpha^2}-\alpha}\,, \label{eq:12}
\end{equation}
Then the decay characteristic length $\xi_{f1}$ and the
oscillation period $\xi_{f2}$,  may be written as
\begin{equation}
    \xi_{f1}=\xi_f / k_1\,, \quad
    \xi_{f2}=\xi_f / k_2\,. \label{eq:13}
\end{equation}
%
The ratio of the characteristic lengths $\xi_{f1} / \xi_{f2} < 1$
clearly shows that magnetic scattering decreases the decay length
and increases the oscillation period \cite{Faure-prb06}.

\subsection{Even mode ($0-$phase)}

In the limit $|a| \ll 1$ we obtain from the even solution
(\ref{eq:10},\ref{eq:11}) the expansion of $F$ in powers of the
amplitude $a$:
\begin{eqnarray}
    F(s)&\simeq& f \cosh(k s) - f^3 g_0(s) + o(f^4) \,, \label{eq:14} \\
    g_0(s)&=& \frac{1}{8 k^2}
     \left[ 4 k s (\alpha + i/4) \sinh(k s) \right.\nonumber \\
        &&\left. +(\alpha + 3i/4) \cosh(3 k s)\right]\,. \nonumber
\end{eqnarray}
Using the first boundary condition (\ref{eq:4}) at $s=s_f$ we get
the relation between the amplitude $a$ and Green's function
$F_s=\sin\theta_s$ in superconductor:
\begin{equation} \label{eq:15}
    f = f_0 + f_0^3 \frac{g_0(s_f)}{\cosh(k s_f)}\,,
    \quad f_0=\frac{F_s}{\cosh(k s_f)}\,.
\end{equation}
Substitution of the solution (\ref{eq:14}) to the relations
(\ref{eq:8},\ref{eq:9}) results in the following equation with
respect to the amplitude $F_s$ in the S layers:
\begin{eqnarray} \label{eq:16}
    \left( \omega + 1/\tau_0 \right) F_s =
    \Delta - \frac{1}{2} \left( \Delta F_s^2
    +\varepsilon \Lambda_0 F_s^3 \right)
\end{eqnarray}
where
\begin{equation} \label{eq:17}
    \tau_0^{-1}= \varepsilon\pi T_{c0} k\, \tanh(k s_f)
\end{equation}
is the depairing parameter of even mode and
\begin{eqnarray}\label{eq:18}
    \Lambda_0&=&\pi T_{c0} \left[ \frac{i}{2 k}\tanh(k s_f) - \right. \\
        &&\left. \frac{\alpha+i/4}{k}\frac{\tanh(k s_f)}{\cosh^2(k s_f)}
        -\frac{s_f (\alpha+i/4)}{\cosh^4(k s_f)} \right]\,. \nonumber
\end{eqnarray}
Here the key parameter
$$
    \varepsilon = \frac{\sigma_f}{\sigma_s} \frac{\xi_s^2}{d_s \xi_f}
$$
determines the influence of the proximity effect on the S layers.

\subsection{Odd mode ($\pi-$phase)}

In the limit $|b| \ll 1$ we obtain from the odd solution
(\ref{eq:10},\ref{eq:11}) the following expansion of $F$ in powers
of the amplitude $b$:
\begin{eqnarray}
    F(s)&\simeq& f\sinh(k s) + f^3 g_{\pi}(s) + o(f^4) \,, \label{eq:19} \\
    g_{\pi}(s)&=& \frac{1}{8 k^2}
     \left[ 4 k s (\alpha + i/4) \cosh(k s) \right.\nonumber \\
        &&\left. -(\alpha + 3i/4) \sinh(3 k s)\right]\,. \nonumber
\end{eqnarray}
Using the first boundary condition (\ref{eq:4}) at $s=s_f$ we get
the relation between the amplitude $b$ and Green's function
$F_s=\sin\theta_s$ in superconductor:
\begin{equation} \label{eq:20}
    f = f_\pi - f_\pi^3 \frac{g_{\pi}(s_f)}{\sinh(k s_f)}\,,
    \quad f_\pi=\frac{F_s}{\sinh(k s_f)}\,.
\end{equation}
Substitution of the solution (\ref{eq:19}) to the relations
(\ref{eq:8},\ref{eq:9}) results in:
\begin{eqnarray} \label{eq:21}
    \left( \omega + 1/\tau_{\pi} \right) F_s =
    \Delta - \frac{1}{2} \left( \Delta F_s^2
    +\varepsilon \Lambda_{\pi} F_s^3 \right)\,,
\end{eqnarray}
where
\begin{equation} \label{eq:22}
    \tau_{\pi}^{-1}= \varepsilon\pi T_{c0} k\, \coth(k s_f)
\end{equation}
is the depairing parameter of odd mode and
\begin{eqnarray}\label{eq:23}
    \Lambda_{\pi}&=&\pi T_{c0} \left[ \frac{i}{2 k}\coth(k s_f) + \right.  \\
        &&\left. \frac{\alpha+i/4}{k}\frac{\coth(k s_f)}{\sinh^2(k s_f)}
        -\frac{s_f (\alpha+i/4)}{\sinh^4(k s_f)} \right]\,. \nonumber
\end{eqnarray}

\section{The critical temperature of SFS trilayer}\label{CritTemper}

%
\begin{figure}[t]
\includegraphics[width=0.50\textwidth]{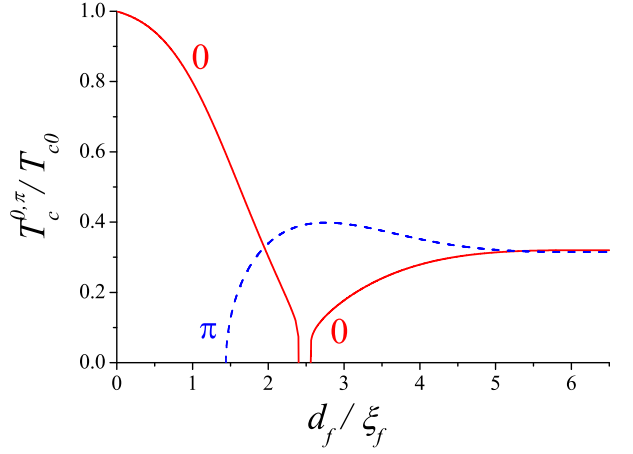}
\caption{(Color online) The typical dependence of the critical
temperature $T_c^{0,\pi}$ on the thickness of F layer $d_f$ for
even mode($0-$phase) (solid red line) and for odd
mode($\pi-$phase) (dashed blue line). Here we choose: $d_s = {\rm
2} \xi_s$; $\sigma_f / \sigma_s = {\rm 0.12}$; $\xi_s / \xi_f =
{\rm 3}$ ($\varepsilon=\mathrm{0.18}$); $h \tau_{s} =
\mathrm{7}$.} \label{Fig:2}
\end{figure}
%
To find the temperature  $T_c^{0,\pi}$ of the second--order
superconducting phase transition, the equations (\ref{eq:16}),
(\ref{eq:21}) should be linearized with respect to the $F_s \ll 1$
\begin{equation} \label{eq:24}
    F_s \simeq F_{s0} = \frac{\Delta}{\omega + 1/\tau_{0,\pi}} \,.
\end{equation}
Substituting Eq.~(\ref{eq:24}) into the self-consistency equation
(\ref{eq:11}) one obtains the equation for the critical
temperature $T_c^{0,\pi}$:
\begin{equation}\label{eq:25}
    \ln\left(\frac{T_c^{0,\pi}}{T_{c0}}\right) = %
        \Psi\left( \frac{1}{2} \right)
          - \mathrm{Re}\,\left[ \Psi\left( \frac{1}{2}
          +\Omega_{0,\pi} \right) \right],
\end{equation}
where $\Psi$ is the digamma function. The depairing parameter
$\Omega_{0,\pi}(T)=1 /2 \pi T \tau_{0,\pi}$
\begin{equation}\label{eq:26}
    \Omega_{0,\pi}=\frac{\varepsilon}{2}\frac{T_{c0}}{T_c^{0,\pi}}
        \begin{cases} k\, \tanh(k s_f), & 0-\mathrm{phase}\\
                      k\, \coth(k s_f), & \pi-\mathrm{phase}
        \end{cases}
\end{equation}
is responsible for the superconductivity destruction in the S
layers due to the proximity effect. Figure~\ref{Fig:2} shows a
typical dependency of the critical temperature $T_c^{0,\pi}$ on
the thickness of F layer $d_f$, obtained from
Eqs.~(\ref{eq:25}),(\ref{eq:26}).
The crossing of the curves $T_c^0(d_f)$ and $T_c^{\pi}(d_f)$
occurs at $d_f^*$, and for $d_f > d_f^*$ the critical temperature
of the $\pi-$phase becomes higher than the critical temperature of
the $0-$phase.

\section{Phase transitions in SFS trilayers} \label{PhaseTrans}

\begin{figure}[t]
\includegraphics[width=0.5\textwidth]{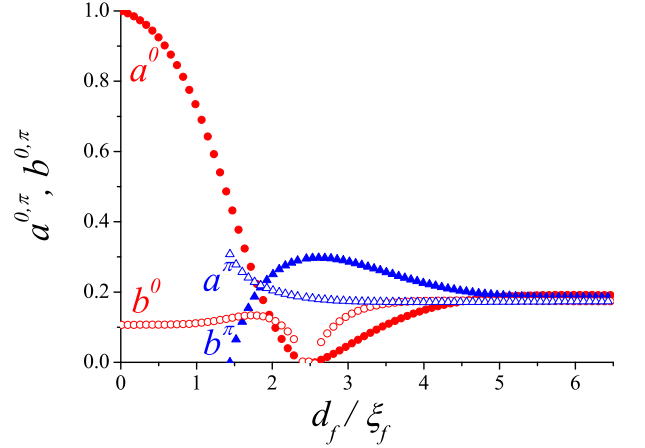}
\caption{(Color online) The typical dependence of the coefficients
of the Ginzburg--Landau expansion $a^{0,\pi}$, $b^{0,\pi}$ on the
thickness of F layer $d_f$: $a^{0}$ ($b^{0}$) -- closed (open) red
circles; $a^{\pi}$ ($b^{\pi}$) -- closed (open) blue triangles.
Here we choose the parameters for Fig.~\ref{Fig:2}.} \label{Fig:3}
%
\end{figure}

To obtain the Ginzburg--Landau (GL) expansion for the $0-$ and
$\pi-$states near the critical temperature $T_c^{0,\pi}$ we can
use the equations (\ref{eq:16}),(\ref{eq:21}).
Assuming $|F_s| \ll 1$ and using $F_{s0}$
(\ref{eq:24}) as a zero-order approximation we find the solution
of Eqs.~(\ref{eq:16}),(\ref{eq:21}) within the first-order
perturbation theory: 
\begin{equation}\label{eq:27}
    F_{s} = \sin\theta_s \simeq F_{s0} -
    \frac{F_{s0}^3}{2} \left[ 1 + \frac{\varepsilon \Lambda_{0,\pi}}
    {\omega + 1/\tau_{0,\pi}} \right]\,.
\end{equation}
Substitution of (\ref{eq:27}) into the self--consistency equation
(\ref{eq:6}) one obtains a dependence of the superconducting gap
$\Delta$ on the temperature $T = T_c^{0,\pi} - \delta T$
\begin{equation}\label{eq:28}
    -a^{0,\pi} \frac{\delta T}{T_c^{0,\pi}} + b^{0,\pi} \Delta^2 = 0 \,,
\end{equation}
where the coefficients $a^{0,\pi}$ and $b^{0,\pi}$ are determined
by the following expressions:
\begin{eqnarray}
    a^{0,\pi}&=&1 - \mathrm{Re} \left[ \Omega_{0,\pi}
        \Psi^{(1)}(1/2+\Omega_{0,\pi}) \right]\,,      \label{eq:29} \\
    b^{0,\pi}&=&\frac{-1}{(4 \pi T_c^{0,\pi})^2}\, \mathrm{Re} \left[
        \Psi^{(2)}(1/2+\Omega_{0,\pi})  \right.    \label{eq:30} \\
        && \left. -\frac{\varepsilon \Lambda_{0,\pi}}
        {6 \pi T_c^{0,\pi}} \Psi^{(3)}(1/2+\Omega_{0,\pi})
        \right]\,,                              \nonumber
\end{eqnarray}
where $\Psi^{(n)}(z) = d^n \Psi(z) / d z^n$ . Naturally all
parameters $a^{0,\pi},b^{0,\pi}$ and $T_{c}^{0,\pi}$ are different
for $0$ or $\pi$ states. Figure~\ref{Fig:3} shows a typical
dependence of the coefficients of the Ginzburg--Landau expansion
$a^{0,\pi}$, $b^{0,\pi}$ on the thickness of F layer $d_f$,
obtained from Eqs.~(\ref{eq:29}),(\ref{eq:30}).
%
%

%
%
\begin{figure*}
\includegraphics[width=0.45\textwidth]{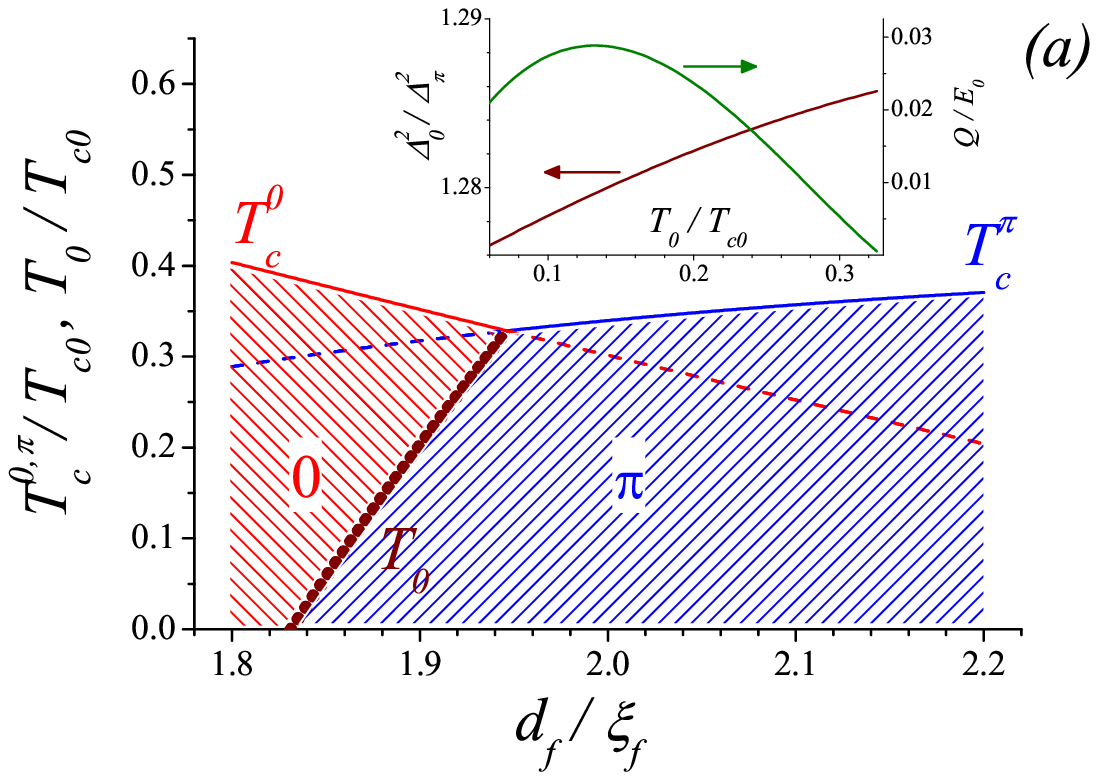}
\includegraphics[width=0.45\textwidth]{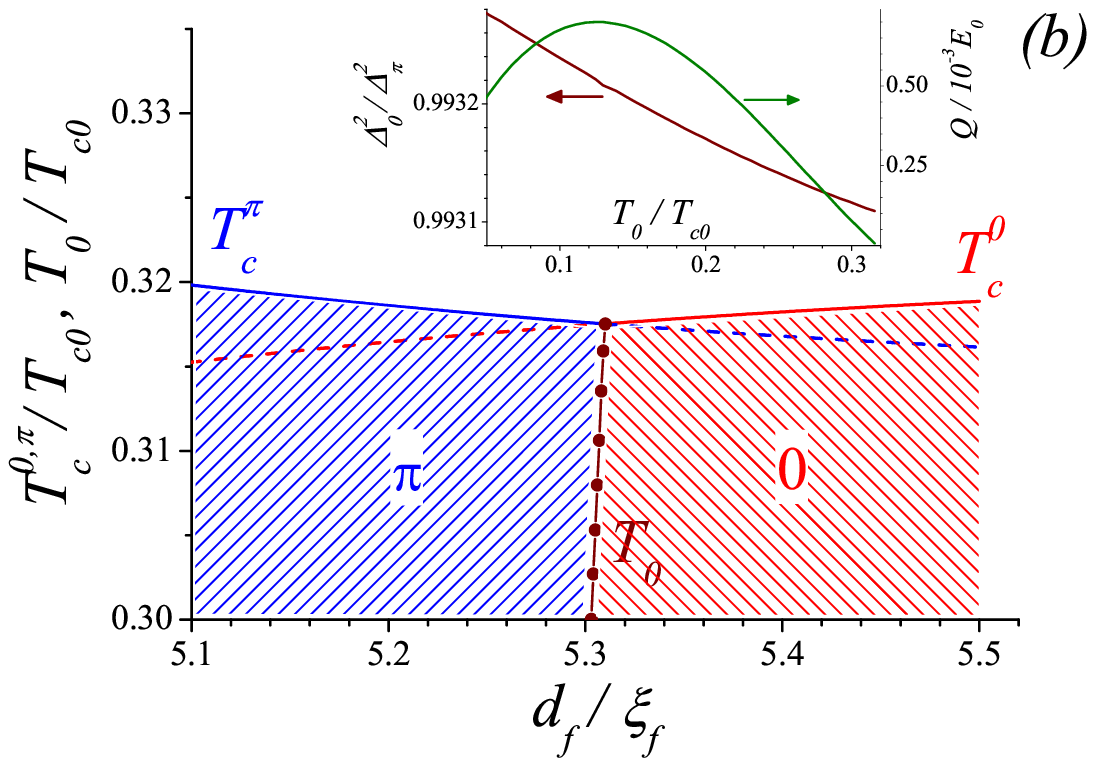}
\caption{The ($T,\,d_{f}$) phase diagram for the SFS trilayers in
the vicinity of crossing ($d_f \approx d_f^*$) of the curves
$T_c^0(d_f)$ and $T_c^{\pi}(d_f)$: (a) $d_f^* \simeq \mathrm{1.95}
\xi_f$; (b) $d_f^* \simeq \mathrm{5.31} \xi_f$ . At temperature
$T_{0}$ the first-order transition between the $0-$
and $\pi-$ states takes place. Here we choose the parameters for Fig.~\ref{Fig:2}.%
The insert gives the dependence of the value of the superconducting gap jump $\Delta_{0}^{2}(T_{0})/\Delta_{\pi}%
^{2}(T_{0})$ (\ref{eq:35}) and the latent heat $Q$ (\ref{eq:34q})
on the transition temperature $T_0$.}\label{Fig:4}
%
\end{figure*}
%
%
The equilibrium value of superconducting gap
\begin{equation}\label{eq:31}
    \Delta_{0,\pi}^2 = \frac{a^{0,\pi}}{b^{0,\pi}}
        \frac{T_c^{0,\pi} - T}{T_c^{0,\pi}}
\end{equation}
corresponds to the extremum of the standard Ginzburg-Landau
functional
\begin{equation}\label{eq:32}
    F_{GL}^{0,\pi}(T) = E_0 \left[ a^{0,\pi} \frac{T - T_c^{0,\pi}}{T_c^{0,\pi}} \Delta^2
    + \frac{b^{0,\pi}}{2}\, \Delta^4 \right] \,,
\end{equation}
where the characteristic energy $E_0 = N(0) S_J d\, T_{c0}^2$ is
determined by the total electron density of states $N(0)$, the
cross section area $S_J$ of the junction, the total thickness of
the trilayer $d = 2 d_s + d_f$ and the critical temperature
$T_{c0}$. The value of the superconducting order parameter
$\Delta$, the temperatures $T$, $T_c^{0,\pi}$ are assumed to be
measured in the units of $T_{c0}$. The functional (\ref{eq:32})
provides us the complete description of the ground states of SFS
junction near the critical temperature $T_{c}^{0,\pi}$. The
equilibrium energy of the system $E^{0,\pi}(T) =
F_{GL}^{0,\pi}(\Delta_{0,\pi})$ is
\begin{equation}\label{eq:33}
    E^{0,\pi}(T) = - E_0 \frac{\left[ a^{0,\pi}\,
    (T_c^{0,\pi}-T)/T_c^{0,\pi} \right]^2}{2 b^{0,\pi}} \,.
\end{equation}
The condition of the first order transition between $0$ and $\pi$
states is $F^{0}(\Delta_{0})=F^{\pi}(\Delta_{\pi}),$ and then the
temperature of this transition $T_{0}$ is determined by:
\begin{equation}\label{eq:34}
    \frac{T_c^0 - T_0}{T_c^{\pi} - T_0}
    = \frac{a^{\pi}}{a^{0}} \sqrt{\frac{b^0}{b^{\pi}}}\,.
\end{equation}
At $d_{f} \lesssim d_{f}^{\ast}$, first the transition from normal
state to the superconducting $0$ state occurs, but further
decrease of the temperature provokes the transition from $0$ to
$\pi$ state. The correspondent ($T,\,d_{f}$) phase diagram for the
SFS trilayers is shown in Fig.~\ref{Fig:4}.

\begin{figure}[t]
\includegraphics[width=0.33\textwidth]{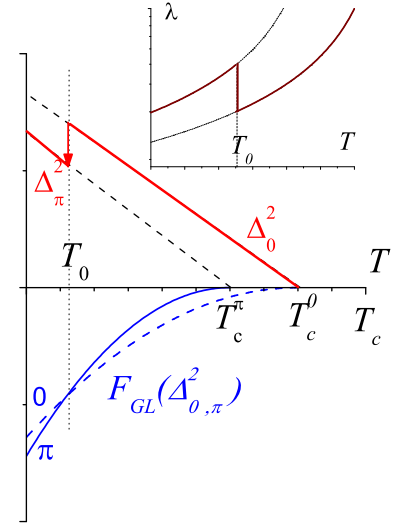}
\caption{(Color online) Schematic dependence of the gap
$\Delta_{0,\pi}^{2}$ (\ref{eq:31}) and the Ginzburg--Landau energy $F_{GL}%
(\Delta_{0,\pi})$ (\ref{eq:33}) on the temperature $T$:
$T_{c}^{\pi} < T_{c}^{0}$ and $(a^{2}/b)^{\pi} > (a^{2}/b)^{0}$ .
The inset gives the schematic temperature
dependence of the penetration depth $\lambda \sim 1 / \Delta_{0,\pi}$.}%
\label{Fig:5}%
\end{figure}

The $0-\pi$ transition is accompanied by a discontinuity in the
entropy $S^{0,\pi}(T) = - \left[\, \partial E^{0,\pi}(T) /
\partial T\, \right]$ at the temperature $T_0$:
\begin{equation}\label{eq-sfs30}
    \frac{S^{\pi}(T_0)}{S^{0}(T_0)} = \frac{T_c^0 - T_0}{T_c^{\pi} - T_0} \,.
\end{equation}
Then the latent heat at the first order $0-\pi$ transition is
\begin{equation}\label{eq:34q}
    Q = \pm T_0 \left[\,S^{\pi}(T_0)-S^{0}(T_0)\,\right] > 0
\end{equation}
for $0 \rightleftarrows \pi$ transitions, respectively.
Simultaneously, at the transition temperature $T_0$ the
superconducting order parameter jumps from $\Delta_{0}$ to
$\Delta_{\pi}$ or vice versa. The ratio of values $\Delta_{0}$ and
$\Delta_{\pi}$ is given by the expression:
\begin{equation}\label{eq:35}
    \Delta_{\pi}^{2}(T_{0})=\frac{a^{\pi}}{b^{\pi}}\left(
        \frac{T_{c}^{\pi}-T_{0}}{T_{c}^{0}}\right)
        =\Delta_{0}^{2}(T_{0})\sqrt {\frac{b^{0}}{b^{\pi}}}\,.
\end{equation}
The inserts in Fig.~\ref{Fig:4} show the dependence
of the ratio $\Delta_{0}^{2}(T_{0})/\Delta_{\pi}%
^{2}(T_{0})$ (\ref{eq:35}) and the latent heat $Q(T_0)$
(\ref{eq:34q}) on the transition temperature $T_0$. Certainly, the
jump of the superconducting gap provokes the jump of the London
penetration depth $1/\lambda \sim \Delta$. Figure~\ref{Fig:5}
shows schematically the temperature dependence of the equilibrium
gap $\Delta_{0,\pi}^2$ (\ref{eq:31}) and the Ginzburg--Landau
energy $E^{0,\pi}(T)$ (\ref{eq:33}) in the vicinity of the $0 \to
\pi$ transition for the case $T_c^{\pi} < T_c^{0}$ and $b^{\pi} >
b^0$. We readily see that the superconducting order parameter
decrease results in the positive jump of the London penetration
depth.  At all reasonable parameters $h,$ $\tau_{s}^{-1},\xi_{f}$
we obtain namely this scenario, which is indeed realized on the
experiment \cite{Pompeo-Samokhvalov-PRB14}.

\section{The current--phase relation}\label{CurrPhaseRelat}

Let us now discuss the peculiarities of the Josephson effect in
the SFS trilayers at the first--order transition from $0$ to $\pi$
state. The theory describing how the $0$ state is transformed into
the $\pi$ state for diffusive SFS junctions was developed in
Ref.~\onlinecite{Buzdin-PRB05_0-pi-trans} for rigid boundary
conditions \cite{Golubov-RMP04} at SF interface. It was shown that
the critical current does not vanish at the transition and is
determined by the second--harmonic term in the current--phase
relation
\begin{equation}\label{eq:42}
    I(\varphi) = I_1 \sin\varphi + I_2 \sin 2 \varphi\,,
\end{equation}
which corresponds to the following phase--dependent contribution
to the free energy of the junction:
\begin{equation}\label{eq:43}
    E_J(\varphi) = \frac{\Phi_0}{2\pi c} \left[
        - I_1 \left(1-\cos\varphi\right)
        - \frac{I_2}{2} \left(1-\cos 2\varphi\right)
        \right]\,,
\end{equation}
where $\Phi_0=\pi \hbar c / e$ is the flux quantum. Since the
amplitudes of the harmonics $I_{1}$ and $I_{2}$ depend strongly on
the superconducting order parameter $\Delta$ in S layers (see
Appendix~\ref{CPR-Rigid} for details), the current--phase relation
of the junction $I(\varphi)$ seems to be very sensitive both to a
suppression of superconductivity in thin S layers and to jumps of
$\Delta$ during the transitions between $0$ and $\pi$ states of
the SFS trilayers.
%
\begin{figure*}[ptb]
\includegraphics[width=0.47\textwidth]{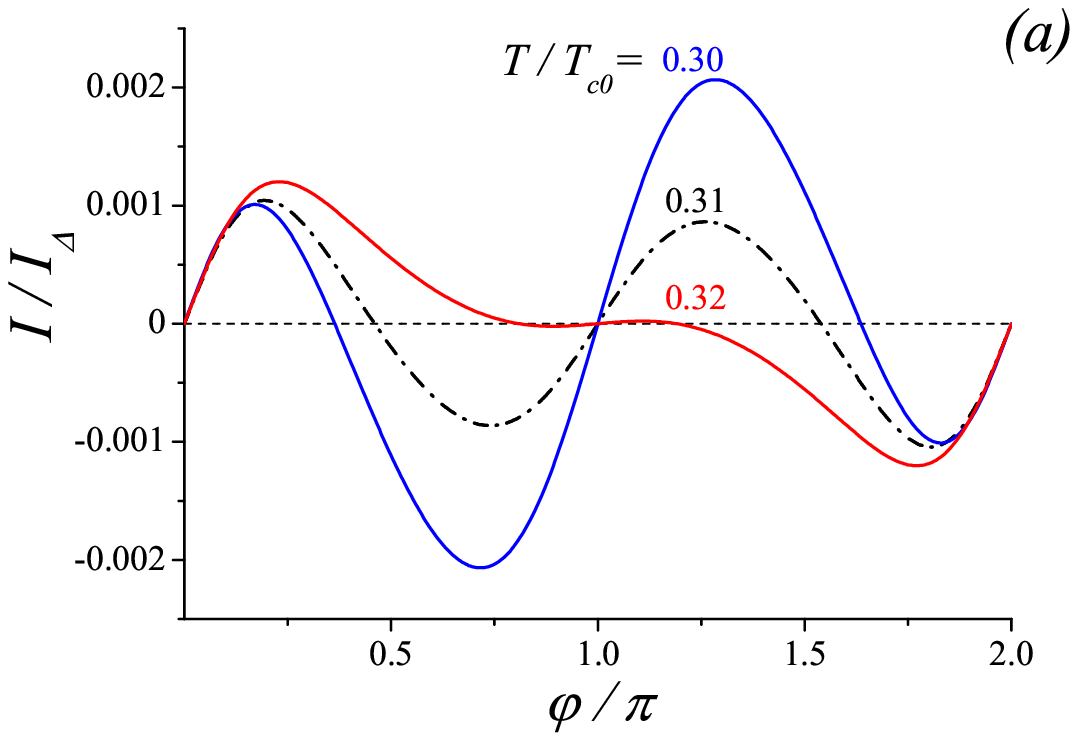}
\includegraphics[width=0.47\textwidth]{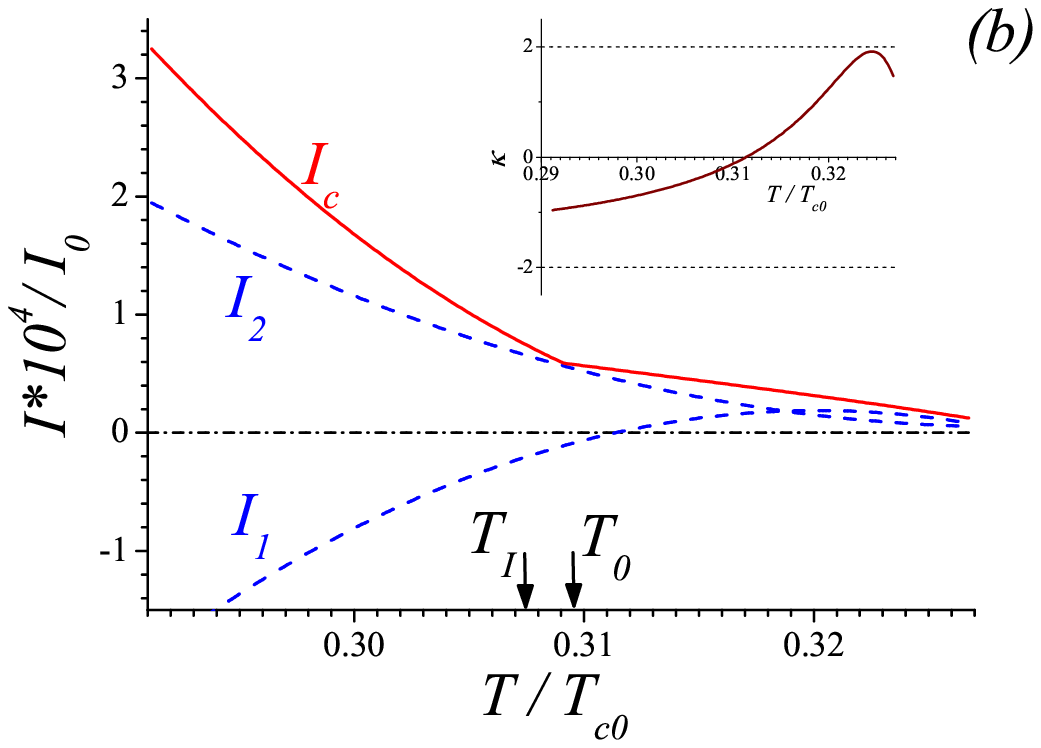}
\caption{(Color online) (a) Current--phase relation of SFS
junction $I(\varphi)$ (\ref{eq:41}) for several values of the
temperature $T / T_{c0}= \mathrm{0.30,\, 0.31,\, 0.32}$ in the
vicinity of the $0-\pi$ transition for $d_f = \mathrm{1.94} \xi_f$
($T_c^0 / T_{c0} \simeq \mathrm{0.332}$, $T_c^\pi / T_{c0} \simeq
\mathrm{0.327}$, $T_c / T_{c0} \simeq \mathrm{0.329}$). The case
$T = T_0 \simeq \mathrm{0.311}\,T_{c0}$ is shown by the
dash-dotted curve ($I_\Delta = I_0\, \Delta_0^2(T)$, where
$\Delta_0(T) = \mathrm{1.764}\,T_c\, \tanh(1.74 \sqrt{T_c/T-1}$ is
BCS superconducting gap for the temperature $T$). (b) Dependence
of the critical current $I_c(T) = \mathrm{max}|I(T,\,\varphi)|$
(red solid line) and the amplitudes of harmonics $I_1$, $I_2$
(blue dashed lines) on temperature $T$. The insert gives the
temperature dependence of a relative amplitude of the first
harmonic $\kappa=I_1 / I_2$. Here we choose the parameters of
Fig.~\ref{Fig:3}: $d_s = {\rm 2} \xi_s$; $\sigma_f / \sigma_s =
0.12$; $\xi_s / \xi_f = {\rm 3}$ ($d_f^* \simeq 1.946 \xi_f$,
$\varepsilon = 0.18$) and $h=10 T_{c0}$, $h \tau_s = 7$.
}\label{Fig:6}
\end{figure*}
%

To consider the Josephson coupling in the SFS trilayers in the
presence of first--order $0 - \pi$ transition, the
Ginzburg--Landau functional (\ref{eq:32}) should be generalized
for arbitrary phase difference $\varphi$ between the
superconducting order parameters in the S layers. For two coupled
S layers with the order parameter $\Delta_{1,2} = \Delta \exp(\pm
i \varphi /2)$ the Ginzburg--Landau expansion $F_{GL}^\varphi$ of
the free energy includes the mixing quadratic term $\Delta_1
\Delta_2^*+\Delta_1^* \Delta_2 \sim \Delta^2 \cos\varphi$. The
mixing terms of fourth order have the form
$(|\Delta_1|^2+|\Delta_2|^2) (\Delta_1 \Delta_2^*+\Delta_1^*
\Delta_2) \sim \Delta^4 \cos\varphi$ and $\Delta_1^2
(\Delta_2^*)^2 + (\Delta_1^*)^2 \Delta_2^2 \sim \Delta^4 \cos
2\varphi$, and provide both $\varphi$ and $2\varphi$ periodicity
of the function $F_{GL}^\varphi$. Therefore, in general, the
Ginzburg--Landau expansion can be written as
\begin{equation}\label{eq:36}
    F_{GL}^\varphi (T, \varphi) / E_0 =  -a_\varphi(T) \Delta^2
        + \frac{b_\varphi}{2} \Delta^4\,,
\end{equation}
which incorporates the phase--dependent contribution via the
coefficients
\begin{eqnarray}
    &&a_\varphi (T) = \gamma_1 \cos\varphi +
         \frac{T_c-T}{T_c^*} (1 + \gamma_2 \cos\varphi)\,,  \nonumber \\
    &&b_\varphi =  \beta_0 + \beta_1 \cos\varphi
            + \beta_2 \cos(2\varphi) \, . \nonumber
\end{eqnarray}
For convenience the superconducting order parameter $\Delta$,
temperatures $T$, $T_c$, $T_c^*$ and the exchange field $h$ are
measured in units of $T_{c0}$. To find the parameters $T_c$,
$T_c^*$, $\gamma_{1,2}$, $\beta_{0,1,2}$ we take into account that
the generalized functional (\ref{eq:36}) reduces to the expression
(\ref{eq:32}) if $\varphi = 0,\,\pi$:
\begin{equation}\label{eq:37}
    F_{GL}^\varphi (T, 0) = F_{GL}^0(T)\,, \quad
    F_{GL}^\varphi (T, \pi) = F_{GL}^{\pi}(T)\,.
\end{equation}
As a result for $T \le \min (T_c^0\,,T_c^{\pi})$ the coefficients
$T_c$, $T_c^*$, $\gamma_{1,2}$, $\beta_{0,1,2}$ can be expressed
as
\begin{eqnarray}
    &&T_c^* = \frac{2}{a^0 / T_c^0 + a^\pi / T_c^\pi}\,, \quad
      T_c = \frac{T_c^*}{2} \left(a^0 + a^\pi\right)\,, \nonumber \\
    &&\gamma_1 = \frac{a^0 - a^\pi}{2} - \frac{T_c}{T_c^*}\,\gamma_2\,, \quad
      \gamma_2 = \frac{T_c^*}{2} \left(\frac{a^0}{T_c^0} -
      \frac{a^\pi}{T_c^\pi}\right)\,,\nonumber \\
    &&\beta_1 = \left( b^0 - b^\pi \right)\, / 2 \,, \quad
      \beta_0 + \beta_2 = \left( b^0 + b^\pi \right)\, / 2 \,. \nonumber
\end{eqnarray}
The coefficient $\beta_2$ in expansion (\ref{eq:36}) can be
determined from the current-phase relation for rigid boundary
conditions (see Appendix~\ref{CPR-Rigid}):
\begin{eqnarray}
    \beta_2 &=& -\frac{\pi}{192} \frac{h}{T_c^3} \frac{\xi_f}{d}\,
        \mathrm{Im}\left\{ \frac{1}{k \sinh^2\delta} \times \right. \label{eq:38} \\
    &&\left.\left[ \frac{\delta}{2} - i \frac{\alpha + i /4}{\sinh\delta}
            \left(\cosh\delta - \frac{\delta}{\sinh\delta}\right)
            \right]\right\}\,,\nonumber
\end{eqnarray}
where $\delta = 2 s_f k = d_f/\xi_{f1} + i d_f/\xi_{f2}$.
Substitution of the equilibrium value of the order parameter
\begin{equation}\label{eq:39}
    \Delta_\varphi^2(T) = a_\varphi(T) / b_\varphi\,.
\end{equation}
into the expression (\ref{eq:36}) provides the temperature
dependence of the free energy $E(T,\,\varphi )$ of the SFS
trilayers for an arbitrary phase difference $\varphi$
\begin{equation}\label{eq:40}
    E(T,\,\varphi) / E_0 = - a_\varphi^2(T) / 2 b_\varphi\,.
\end{equation}
which results in the following current--phase relation $I(\varphi)
= (2e /\hbar)\, \partial E / \partial\varphi$:
\begin{eqnarray}
    I(T,\,\varphi) &=& I_0\,\sin\varphi\,
    \frac{a_\varphi(T)}{b_\varphi} \times \label{eq:41} \\
    &&\left[\,\gamma_1+\gamma_2\frac{T_c-T}{T_c^*}
        -\frac{a_\varphi(T)}{2 b_\varphi}(\beta_1+4\beta_2\cos\varphi)\,\right]\,, \nonumber
\end{eqnarray}
where $I_0=2\pi c\, E_0 / \Phi_0$. The current--phase relation
$I(\varphi)$ (\ref{eq:41}) is shown in Fig.~\ref{Fig:6}a for fixed
thickness of the barrier $d_f = \mathrm{1.94} \xi_f$ in the
vicinity of the $0-\pi$ transition and several values of the
temperature $T$. For chosen parameters of SFS trilayers (see
Fig.~\ref{Fig:3}) the first $0$-$\pi$ transition takes place at
$d_f \lesssim 1.95 \xi_f$. Corresponding value of the coefficient
$\beta_2 \simeq \mathrm{-0.013}$ gives a rather large value of the
second--harmonic term ($|I_2| \sim |I_1|$), which dominates near
the transition. Due to the strong contribution of higher harmonics
the current--phase relation is rather anharmonic, and the maximal
value of $|I(T,\,\varphi)|$ occurs at $\varphi \ne \pi/2$.
Figure~\ref{Fig:6}b shows the temperature dependence of the
critical current $I_c(T) = \mathrm{max}_\varphi\,|I(T,\,\varphi)|$
near the temperature $T_0$. The characteristic multimode
anharmonicity of the current-phase relation results in the
disappearance of the typical nonmonotonic temperature dependence
of the critical current in a vicinity of the $0-\pi$ transition.
We have obtained the positive amplitude of second harmonic $I_2 >
0$, which means that it occurs discontinuously by a jump between
$0$ and $\pi$ phase states at the transition point $T_I$, where
the critical current $I_c(T)$ formally changes it sign. The shift
of the temperature $T_I$ with respect to $T_0$ depends on the
higher harmonics contribution.

%
\begin{figure*}[ptb]
\includegraphics[width=0.47\textwidth]{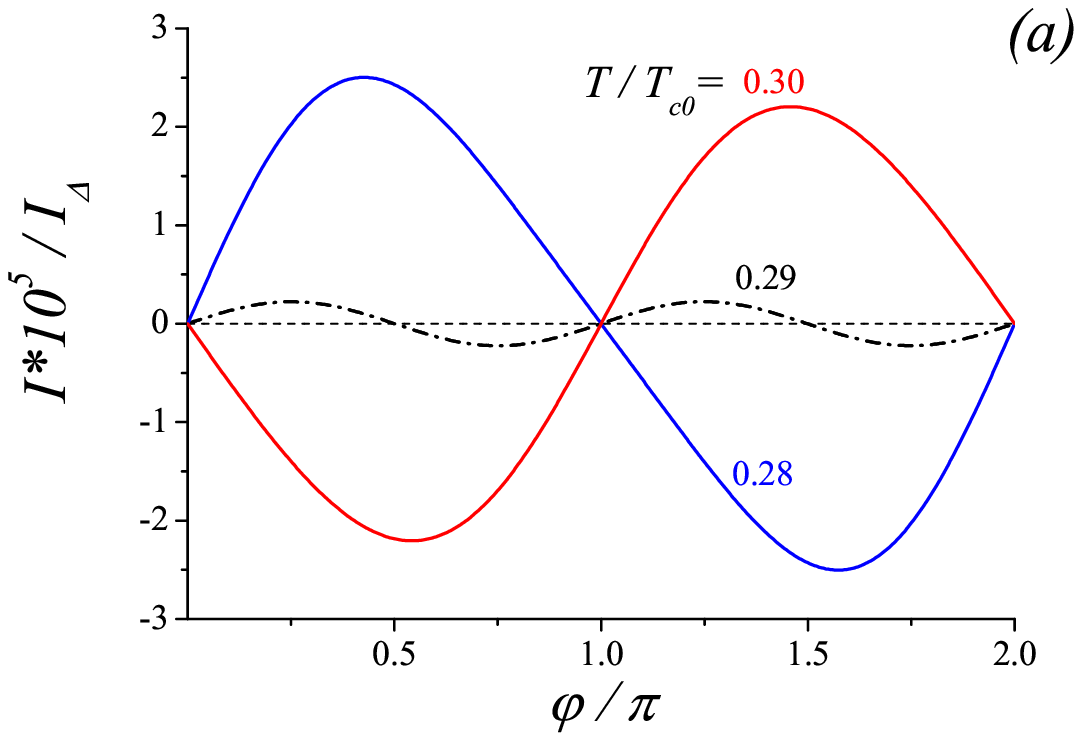}
\includegraphics[width=0.47\textwidth]{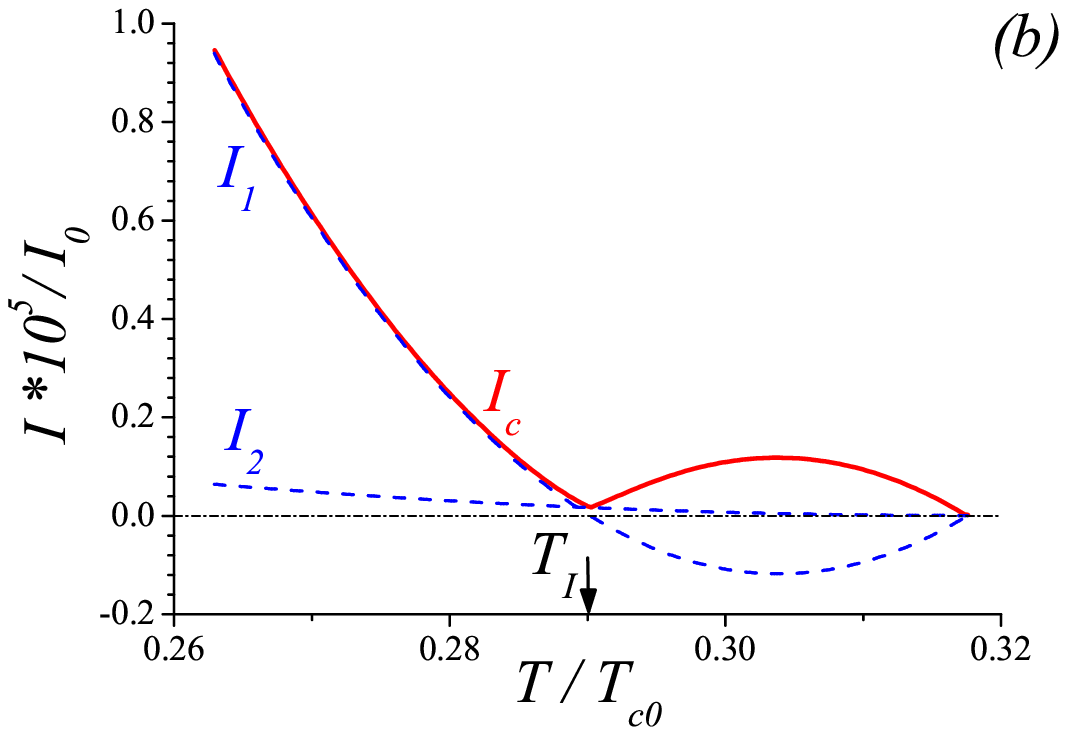}
\caption{(Color online) (a) Current--phase relation of SFS
junction $I(\varphi)$ (\ref{eq:41}) for several values of the
temperature $T / T_{c0}= \mathrm{0.28,\, 0.29,\, 0.30}$ in the
vicinity of the second $0-\pi$ transition for $d_f = \mathrm{5.3}
\xi_f$ ($T_c^0 / T_{c0} \simeq \mathrm{0.317}$, $T_c^\pi / T_{c0}
\simeq \mathrm{0.318}$, $T_c / T_{c0} \simeq \mathrm{0.3175}$ ).
The case $T = T_0 \simeq \mathrm{0.29}\,T_{c0}$ is shown by the
dash-dotted curve ($I_\Delta = I_0\, \Delta_0^2(T)$, where
$\Delta_0(T) = \mathrm{1.764}\,T_c\, \tanh(1.74 \sqrt{T_c/T-1}$ is
BCS superconducting gap for the temperature $T$). (b) Dependence
of the critical current $I_c(T) = \mathrm{max}|I(T,\,\varphi)|$
(red solid line) and the amplitudes of harmonics $I_1$, $I_2$
(blue dashed lines) on temperature $T$. Here we choose the
parameters of Fig.~\ref{Fig:3}: $d_s = {\rm 2} \xi_s$; $\sigma_f /
\sigma_s = 0.12$; $\xi_s / \xi_f = {\rm 3}$ ($d_f^* \simeq 5.31
\xi_f$, $\varepsilon = 0.18$) and $h=10 T_{c0}$, $h \tau_s = 7$.
}\label{Fig:7}
\end{figure*}
%
For comparison figure~\ref{Fig:7} shows the current--phase
relations $I(\varphi)$ (\ref{eq:41}) and the temperature
dependence of the critical current $I_c(T)$ in a vicinity of the
second $\pi-0$ transition (see Fig.~\ref{Fig:3}) which takes place
at $d_f \lesssim 5.31 \xi_f$. Corresponding value of the
coefficient $\beta_2 \simeq -0.18\, 10^{-4}$ gives a small value
of the second--harmonic term ($|I_2| \ll |I_1|$), and the the
current--phase relation is quite harmonic except the case $T
\approx T_0 \simeq T_I$. The SFS junction reveals the typical
nonmonotonic behavior of the critical current $I_c$ as a function
of the temperature $T$, and the position of the cusp $T_I$
naturally coincides with the temperature $T_0$.

If we restrict our consideration to the two harmonics approach
(\ref{eq:42}), one will be able to determine the amplitude of both
harmonics $I_1$ and $I_2$ via critical current $I_c$ measurements,
as it has been proposed in
Ref.~\onlinecite{Goldobin-Buzdin-prb07}. Since a relative
amplitude of the first harmonic $\kappa = I_1 / I_2$ is small
($|\kappa| < 2$) in the vicinity of $0-\pi$ transition (see the
insert in Fig.~\ref{Fig:6}b), the system has two stable states
$\varphi=0$ and $\varphi=\pi$ at $I=0$. To "depin"\ the Josephson
phase from the low energy $0$ ($\pi$) state or from the high
energy $\pi$ ($0$) for $I_1 > 0$ ($I_1 < 0$), respectively, the
critical current
\begin{eqnarray}
    I_{c\pm}(\kappa) &=& \frac{I_2}{32}
        \left(\sqrt{\kappa^2+32} \pm 3 |\kappa| \right)^{3/2}
        \label{eq:70} \\
        &\times&\left(\sqrt{\kappa^2+32} \mp |\kappa|
        \right)^{1/2}\, \nonumber
\end{eqnarray}
should be applied \cite{Goldobin-Buzdin-prb07}. During switching
from the voltage state to the zero resistance state at $T \approx
T_0$, the phase may stick in the high energy state with a
probability close to $50\%$, and the smaller current $I_{c-}$
becomes measurable in the SFS structure with a low damping
\cite{Sickinger-Goldobin-prl12}. Then for $|\kappa| \ll 1$ near
$0-\pi$ transition the amplitude of the harmonics $I_{1,2}$ is
determined by the relations
\begin{equation}\label{eq:71}
    |I_1| \simeq \left( I_{c+} - I_{c-} \right) / \sqrt{2}\,, \qquad
    I_2 \simeq \left( I_{c+} + I_{c-} \right) / 2 \,.
\end{equation}

\begin{figure}[ptb]
\includegraphics[width=0.3\textwidth]{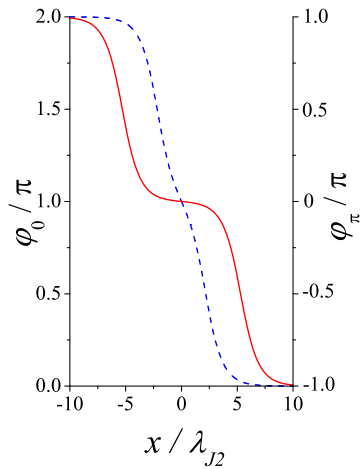}
\caption{(Color online) The shape of the Josephson vortex
$\varphi_{0,\pi}(x)$ (\ref{eq:72}),(\ref{eq:74}) for two values of
the relative amplitude of the first harmonic $\kappa = I_1 / I_2$
in the vicinity of the $0-\pi$ transition: blue dashed line --
$|\kappa|=\mathrm{0.05}$; red solid line --
$|\kappa|=\mathrm{10^{-4}}$. }\label{Fig:8}
\end{figure}

The structure of the Josephson vortex which may exist in a long
Josephson junction with a large second harmonic ($|\kappa| < 2$)
is rather peculiar and is described by the following expressions
\cite{Buzdin-PRB05_0-pi-trans}
\begin{equation}
    \varphi_0 = \left\{
       \begin{array}{cc}
          2\pi-\varphi_\kappa(x/\lambda_{J2}) , & x < 0 \label{eq:72} \\
          \varphi_\kappa(x/\lambda_{J2}) , & x > 0
       \end{array} \right.\,, \quad \varphi_\pi = \varphi_0 - \pi
\end{equation}
for $I_1 > 0$ ($I_1 < 0$), respectively, where
\begin{equation}\label{eq:74}
    \varphi_\kappa(\chi) = \mathrm{arccos} \left(1 -
        \frac{2(1+|\kappa|)}
             { 1 + |\kappa| \mathrm{cosh}^2(\chi \sqrt{1+|\kappa|})}
        \right)
\end{equation}
and the Josephson length $\lambda_{J2}^{-2} = c \Phi_0 S / 8\pi^2
I_2 t$ depends on the current density $I_2 / S$ and the effective
junction thickness $t$. The change of the form of the Josephson
vortex in the vicinity of the $0-\pi$ transition is shown in
Fig.~\ref{Fig:8}.
%
%

\section{Single--Junction loop}\label{SingleJuncLoop}

\begin{figure}[ptb]
\includegraphics[width=0.47\textwidth]{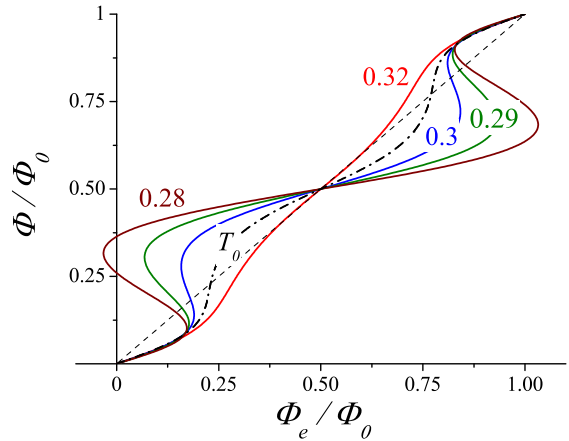}
\caption{(Color online) Magnetic flux through the single--junction
loop $\Phi$ as a function of the external flux  $\Phi_e$ for the
normalized inductance $L_I = \mathrm{250}$ and several values of
the temperature $T / T_{c0}= \mathrm{0.28,\, 0.29,\, 0.3,\,
0.31,\, 0.32}$ in the vicinity of the $0-\pi$ transition for $d_f
= \mathrm{1.94} \xi_f$. The case for $T = T_0 \simeq
\mathrm{0.31}\,T_{c0}$ is shown by the dash-dotted curve. Here we
choose the parameters of Fig.~\ref{Fig:6} and $L_I =
\mathrm{500}$.}\label{Fig:9}
\end{figure}

Let us consider a small superconducting loop with an inductance
$L$ interrupted by the single SFS junction (rf SQUID). We assume
that the junction is described by the current--phase relation
(\ref{eq:41}) and choose the parameters in a narrow region near
the $0 - \pi$ transition. The ground state of the circuit is
determined by minimizing the SQUID free energy
\begin{equation}\label{eq:60}
    W(T,\,\phi) = E(T,\,\phi)
        + \frac{\Phi_0^2}{8\pi^2 L} \left(\phi - \phi_e \right)^2\,,
\end{equation}
where $\phi_e = 2\pi \Phi_e / \Phi_0$ -- is the normalized
magnetic flux of an external field through the loop
\cite{Barone-Paterno_Joseph}. The total magnetic flux through the
loop $\Phi = (\phi / 2\pi)\,\Phi_0$ is related to the external
flux $\Phi_e$ as
\begin{equation}\label{eq:61}
    \phi_e = \phi + L_I I(T,\,\phi) / I_0\,,
\end{equation}
with the normalized inductance $L_I = 2\pi L\, I_0 / c \Phi_0$. 
Figure~\ref{Fig:9} shows the dependence $\Phi(\Phi_e)$ determined
by Eqs.~(\ref{eq:60}) and (\ref{eq:61}) for several values of the
temperature $T$ near the transition temperature $T_0$.
%
%
New features of $\Phi(\Phi_e)$ dependence appear for $T$ close to
$T_0$ due to the coexistence of stable and metastable $0$ and
$\pi$ states.  Strong anharmonicity of the current-phase relation
for $T \approx T_0 \sim T_c$ results in the coexistence of integer
and half--integer fluxoid configuration in SQUID's and generation
of two flux jumps per one external flux quantum. This behavior is
similar to the magnetic-flux penetration in superconducting loop
with a clean SFS junction at low temperatures $T \ll T_c$
\cite{Radovic-PRB01_0-pi-SQUID}.

\section{Summary}\label{Summary}

We have studied the thermodynamics of diffusive SFS trilayer with
relatively thin S layers through self-consistent solutions of
nonlinear Usadel equations, in the dirty limit. Our results may be
viewed as generalization of those obtained in
Ref.~\onlinecite{Buzdin-PRB05_0-pi-trans}, when the
superconducting electrodes are the rather thin, and the critical
temperature $T_c$ is affected by the ferromagnetic layer. We have
shown that as the temperature $T$ is varied a given SFS junction
can flip from the $0$ state to the $\pi$ state. The resulting
phase transition is first--order, in agreement with the
experiments
\cite{Frolov-Ryazanov-PRB04,Sellier-PRL04,Oboznov-Ryazanov-Buzdin-PRL06}
and is responsible for a jump of the amplitude of the
superconducting order parameter $\Delta$, providing the anomalous
temperature behavior of the effective penetration depth
$\lambda(T)$ observed in
Ref.~\onlinecite{Pompeo-Samokhvalov-PRB14}. Taking the typical
parameters $N(0) \sim \mathrm{10^{33}\,erg^{-1}\, sm^{-3}}$,
$T_{c0} \sim \mathrm{10\, K} \approx \mathrm{10^{-15}\,erg}$ we
may estimate for the $\mathrm{Nb/Pd_{0.84}Ni_{0.16}/Nb}$ trilayer
\cite{Torokhtii-PhysC12,Pompeo-Samokhvalov-PRB14} ($d \simeq
\mathrm{30\,nm}$, $S \sim \mathrm{0.1\, sm^2}$) the latent heat $Q
\sim \mathrm{0.01} E_0 \approx \mathrm{0.3\, pJ}$. Such picojoule
latent heat can be readily observed via standard ac calorimetry
techniques used to measure specific and latent heats in films:
attojoule level results have been reported \cite{Bourgeois-PRL05}.
We see, therefore, that the latent heat $Q$ (\ref{eq:34q})
associated with the first--order $0 - \pi$ transition in SFS
trilayer is quite observable.

We have proposed the general form of the Ginzburg-Landau
functional to describe SFS trilayer for arbitrary phase difference
$\varphi$ between the order parameters in the superconducting
layers. Calculation of the current--phase relation $I(\varphi)$
shows that the ground state of the SFS junction is $0$ or $\pi$,
and the transition between the $0$ and $\pi$ states appears
discontinuous. The current--phase relation strongly deviates from
the simple sinusoidal one due to strong dependence of
superconductivity in thin S layers on the structure of the pair
wave function in the ferromagnetic, even at temperatures $T$ near
the critical value $T_c$. The characteristic anharmonicity of the
current--phase relation results in disappearance of the typical
nonmonotonic temperature dependence of the critical current in a
vicinity of the $0-\pi$ transition. Certainly, the anharmonicity
of the current--phase relation becomes less pronounced for thick
ferromagnetic layer, if $d_f \gg \xi_f$.

We show that coexisting stable and metastable $0$ and $\pi$ states
appear in the vicinity of $0-\pi$ transition. As a consequence,
integer and half-integer fluxoid configurations exist in the
superconducting loop interrupted by the junction. The coexistence
of $0$ and $\pi$ states is manifested as two jumps in the
dependence of enclosed magnetic flux in the loop per period
\cite{Radovic-PRB01_0-pi-SQUID}.

acknowledgments{The authors thank E. Silva, V.V. Ryazanov and A.S.
Mel'nikov for stimulating discussions. This work was supported, in
part, by the NanoSC-COST (Belgium), Action MP1201 , by ANR
(France), Grant MASH and by the Russian Foundation for Basic
Research. One of the authors (A.V.S.) is supported by the Russian
Scientific Foundation Grant No 15-12-10020.}\\

\appendix

\section{The current--phase relation for rigid boundary
conditions}\label{CPR-Rigid}

Let us briefly remind of approach developed in
Ref.~\onlinecite{Buzdin-PRB05_0-pi-trans} to study the problem of
the second--harmonic contribution to the the current--phase
relation (\ref{eq:42}) in the framework of the rigid boundary
conditions. The general expression for the supercurrent through a
SFS junction is given by
\begin{equation}\label{eq:44}
    I_s = - I_0 (T/T_c) \sum\limits_{\omega > 0} \mathrm{Im} \left\{ F^+ F^\prime_s \right\}\,,
\end{equation}
where $F^+(s, h) = F^*(s,-h)$, $I_0 = 4\pi T_c e S N(0) D_f /
\xi_f$, $S$ is the area of cross section of the junction, and
$N(0)$ is the electron density of states per one-spin projection.

The solution of Eqns.~(\ref{eq:1}), (\ref{eq:7}) for $T \lesssim
T_c$ and an arbitrary phase difference $\varphi$ is
\cite{Buzdin-PRB05_0-pi-trans}
\begin{eqnarray}\label{eq:45}
    &&F(s) \simeq a\, \cosh(q s) + b\, \sinh(p s) \\
    &&\quad + \frac{b^2 - a^2}{8 k^2} \left(\alpha + \frac{3 i}{4}\right)
       \left[ a \cosh(3 k s) + b \sinh(3 k s) \right]\,,  \nonumber
\end{eqnarray}
where the complex wave vectors $q$ and $p$ are different from the
wave vector $k$ due to nonlinear effects. Just below $T_c$ the
nonlinear corrections seem to be small and wave vectors $q$ and
$p$ are determined by the relations
\begin{eqnarray}
    &&q \simeq k - \frac{a^2}{2 k}\left(\alpha + i / 4 \right)
        + \frac{b^2}{2 k}\left(\alpha + 5 i / 4 \right) \,, \\      \label{eq:48}
    &&p \simeq k -\frac{a^2}{2 k}\left(\alpha + 5 i / 4 \right)
        - \frac{b^2}{2 k}\left(\alpha + i / 4 \right) \,.           \label{eq:49}
\end{eqnarray}
The function $F^+(s)$ is obtained by replacing $b$ in expression
(\ref{eq:45}) by $-b$. Neglecting the influence of the F layer on
the S layers, one can find the amplitudes $a$ and $b$ from the
rigid boundary conditions assuming that the anomalous Green's
function for the Matsubara frequency $\omega$ at the boundary of
left (right) S layer coincides with the bulk one: $F_s(\omega,\mp
s_f) = \Delta_B\, \mathrm{e}^{\pm i\varphi/2} / \Omega$, where
$\Delta_B$ is the temperature dependent BCS order parameter and
$\Omega = \sqrt{\omega^2 + |\Delta_B|^2}$. Using the boundary
conditions
\begin{equation}\label{eq:50}
    F(\pm s_f) = F_s(\omega,\pm s_f)
\end{equation}
we get from (\ref{eq:45}),(\ref{eq:48}),(\ref{eq:49}) the
following expressions for the amplitudes $a$ and $b$:
\begin{equation}\label{eq:51}
    a = a_0 + a_1\,, \quad b = b_0 + b_1\,,
\end{equation}
\begin{equation}\label{eq:52}
    a_0 = \frac{\Delta_B}{\Omega} \frac{\cos(\varphi/2)}{\cosh(\delta/2)} \,,
    \quad
    b_0 = - i \frac{\Delta_B}{\Omega} \frac{\sin(\varphi/2)}{\sinh(\delta/2)} \,,
\end{equation}
\begin{widetext}
\begin{eqnarray}
    &&a_1 = \frac{-a_0}{8 k^2 \cosh(\delta/2)} \left[ \left( b_0^2-a_0^2 \right)
        \left(\alpha + 3 i /4 \right) \left( \cosh(3\delta/2) + 2\delta \sinh(\delta/2) \right)
        + i \delta \left( a_0^2 + b_0^2 \right) \sinh(\delta/2) \right]\,, \label{eq:53}  \\
    &&b_1 = \frac{-b_0}{8 k^2 \sinh(\delta/2)} \left[ \left( b_0^2-a_0^2 \right)
        \left( \alpha+3 i /4 \right) \left(\sinh(3\delta/2) - 2\delta \cosh(\delta/2) \right)
        + i \delta \left( a_0^2 + b_0^2 \right) \cosh(\delta/2) \right]\,, \label{eq:54}
\end{eqnarray}
where $\delta = \delta_1 + i \delta_2$ and $\delta_{1,2} = d_f /
\xi_{f1,f2}$. Substitution of the solutions (\ref{eq:45}),
(\ref{eq:51}), (\ref{eq:52}), (\ref{eq:53}) to the general
expression for the total Josephson current (\ref{eq:44}) results
in the following expression for the amplitude of the first
harmonic $I1$ of the current--phase relation (\ref{eq:42}):
\begin{equation}
    I_1 = \frac{I_0}{8} \left( \frac{\Delta_B}{T_c} \right)^2 \left[
          \mathrm{Im}\left\{ \frac{i k}{\sinh\delta} \right\}
            -\frac{1}{12} \left(\frac{\Delta_B}{T_c}\right)^2
          \mathrm{Im}\left\{ \frac{i (\alpha+5 i / 4)}{k \sinh\delta}
            - \frac{i (\alpha + i/4) \left( 1 - \delta/\tanh\delta \right) }{k \sinh^3\delta}
                    \right\} \right] \,.   \label{eq:56}
\end{equation}
The second harmonic amplitude is much smaller and described by the
following expression
\begin{equation} \label{eq:57}
    I_2 =\frac{I_0}{192} \left(\frac{\Delta_B}{T_c}\right)^4 \mathrm{Im}\left\{
          \frac{1}{k \sinh^2\delta} \left[ \frac{\delta}{2} - i \frac{\alpha + i /4}{\sinh\delta}
            \left(\cosh\delta - \frac{\delta}{\sinh\delta}\right)
            \right]\right\} \,.
\end{equation}
Actually the Josephson energy (\ref{eq:43}) is the phase dependent
contribution to the Ginzburg-Landau energy (\ref{eq:36}).
\begin{equation}\label{eq:58}
    \frac{1}{2}\, E_0 \left(\frac{\Delta}{T_{c0}} \right)^4 \beta_2 \cos 2\varphi \equiv
    -\frac{\Phi_0 I_2}{4\pi c} \cos 2\varphi \,.
\end{equation}
The last equality determines the coefficient $\beta_2$
(\ref{eq:38}) in the functional (\ref{eq:36}):

\begin{equation}\label{eq:59}
    \beta_2 = -\frac{\pi}{192} \left(\frac{T_{c0}}{T_c}\right)^3 \frac{h}{T_{c0}} \frac{\xi_f}{d}\,
        \mathrm{Im}\left\{ \frac{1}{k \sinh^2\delta}
            \left[ \frac{\delta}{2} - i \frac{\alpha + i /4}{\sinh\delta}
            \left(\cosh\delta - \frac{\delta}{\sinh\delta}\right)
            \right]\right\}\,,
\end{equation}

\end{widetext}

\end{document}